\documentclass[useAMS,usenatbib]{mn2e}

\usepackage{graphicx}
\usepackage{times}
\title[Measuring stellar magnetic fields with ISIS@WHT]{Measuring stellar magnetic fields with the low resolution spectropolarimeter of the William Herschel Telescope}

\author[F. Leone]{F. Leone\\
Universit\`a di Catania, Dipartimento di Fisica e Astronomia -- Sezione Astrofisica, Via S. Sofia 78, I--95123 Catania, Italy E-mail:  Franco.Leone@oact.inaf.it}

\begin{document}

\date{Received To be inserted later, accepted To be inserted later}


\maketitle

\label{firstpage}

\begin{abstract}
Despite the influence of magnetic fields on the structure and evolution of
stars is largely demonstrated from the theoretical point of view,
their observational evidence in non-degenerated stars is still rather scanty
and mainly circumscribed to bright objects (V$<$10). Stellar magnetic fields
are commonly measured on the basis of circular spectropolarimetry at high/middle
resolution across the profile of metal lines. The present sensitivity
of telescopes and spectrographs makes this still an almost prohibitive method
for faint stars. In principle, stellar magnetic fields can be also measured
on the basis of low resolution spectropolarimetry, with very important results
obtained at the  8\,m ESO telescopes with FORS1. The trade off between S/N and
spectral resolution in measuring stellar magnetic fields justify an attempt,
here presented, to perform these measurements at the  4.5 m $William\ Herschel\ Telescope$. HD\,3360, one of the stars with the weakest
known magnetic field, and the magnetic chemically
peculiar stars HD\,10783, HD\,74521 and HD\,201601 have been
observed  with the
spectropolarimeter ISIS in the 3785$-$4480\,\AA\ range.
Measured stellar magnetic fields, from Stokes $I$ and $V$ spectra with
S/N $>$ 600, show an internal error of $\le$50 G selecting the whole
interval and $\le$200 G within a Balmer line. Ripples in the Stokes $V$ spectra of HD\,3360 result in
an instrumental positive magnetic field certainly not larger than 80 G.
\end{abstract}

 
\section{Introduction}
By means of circular spectropolarimetry across metal line profiles,
Babcock \cite{Babcock47} has shown that
is possible to measure the effective magnetic field of stars: $\rm B_{\rm eff}$, i.e. the
line-intensity weighted average over the visible stellar disk of the
line-of-sight component of the magnetic field vector.
For a spectral line at $\lambda_0$, Mathys \cite{Mathys94} related the first order moment $R_V^{(1)}$ of Stokes $V$ profile
and the effective magnetic field: 
\begin{equation}
R_V^{(1)} = \frac{1}{W}\int{\frac{V_c - V_{\lambda}}{I_c}\,(\lambda-\lambda_0)\,\delta\lambda} = \Delta\lambda_z g_{\rm eff} \lambda_0^2 \rm B_{\rm eff} \label{eqRV}
\end{equation}
where $W$ is the equivalent width, $I$ is the intensity profile, the subscript $c$ refers to the continuum,
$g_{\rm eff}$ is the effective Land\'e factor, $\lambda$ wavelengths are expressed in \AA\ and
$\rm{B_{\rm eff}}$ is in Gauss and $\Delta\lambda_z = 4.67\,10^{-13}\rm \AA^{-1}G^{-1}$. 

The sensitivity of the present generation of telescopes and spectrographs
let the ideal very high (R$>$100\,000) resolution circular spectropolarimetry
and even the (R=15\,000) moderate one prohibitive to routinely detect the
magnetic field of faint non-degenerate stars. At my knowledge, BP\,Tau (V=12.3) and DF\,Tau
(V=11.0) are among the few faint stars whose magnetic field geometry has been
inferred from $\rm{B_{\rm eff}}$ measurements \cite{Symington05}.
On the basis of the relation, strictly valid in the weak field hypothesis
and other approximations listed in Mathys \cite{Mathys89}:\begin{equation}
\frac{V}{I} = -g_{\rm eff} \ \Delta \lambda_z \ \lambda^{2}
\frac{1}{I} \frac{{\rm d}I}{{\rm d}\lambda} {\rm{B_{\rm eff}}} \label{eqV}
\end{equation} 
 Angel \& Landstreet \cite{Angel70}
introduced a new method to measure the field of white dwarfs
performing narrow ($\sim$30\AA) band circular
photopolarimetry on the wings of $H_{\gamma}$ Balmer line.

Later on, Bagnulo et al. \cite{Bagnulo02} were successful in measuring stellar
magnetic fields by means of the low resolution spectropolarimeter FORS1 at
the 8\,m {\it Antu} ESO telescope. These authors selected the blue region,
to take advantage of the Balmer lines crowding to their series limit, with the
grism 600B ({\it linear dispersion} = 1.50 \AA/pix) and used the
0.5$\arcsec$ slit (R=1600). Bagnulo and coworkers showed that the effective
magnetic field of V=13 star can be measured with a 150 G error in two hours
(see also Bagnulo et al. \cite{Bagnulo04}).   
A fundamental result for the understanding of the role of magnetic field
in structure and evolution of stars, as testified by the rapid application
of this method from the pre-main sequence objects \cite{Wade07} to evolved
stars \cite{Beuermann07}. 

A measure of $\rm{B_{\rm eff}}$ at low resolution consists of a linear
regression of $V/I$ versus
$-g_{\rm eff} \ \Delta \lambda_z \ \lambda^{2} \ (1/I) \ {\rm d}I/{\rm d}\lambda$,
whose accuracy is proportional to $V/I$ error
 ($\propto S/N$ per pixel) and to the square root of number of points sampling the
spectrum ($\propto 1/\sqrt{linear\, dispersion}$). This trade off between
S/N per pixel and linear dispersion justifies the evaluation of measuring
stellar magnetic fields with the 4.5 m {\it William Herschel Telescope} (WHT)
equipped with the spectropolarimeter ISIS at 
{\it linear dispersion} = 0.23 \AA/pix and R=5000 (slit=0.5$\arcsec$).

A comparison shows that the capability to measure stellar magnetic fields
with ISIS at the WHT and FORS1 as used by Bagnulo and coworkers
at the 8\,m telescopes of the ESO VLT is not as large as instinctively supposed.
The Exposure Time Calculators for a 0.8$\arcsec$ seeing show that FORS1
gives a S/N per pixel about six times larger in the $B$ band. An advantage
that is almost balanced by a linear dispersion seven times larger for ISIS.
Thus, stellar magnetic fields measured with FORS1 are in principle two
times more accurate only. A higher ISIS efficiency in measuring stellar
magnetic fields should derive from its larger spectral resolution because
of a reduced smoothing and blending of Stokes $V$ line profiles. 
Leone et al. \cite{Leone00} have numerically shown that, at a given S/N,
doubling the rotational velocity of a star the error in measuring its effective
magnetic field through eq.\,(\ref{eqRV}) is doubled too. 

 FORS1 has been upgraded in April 2006 with the grism 1200B
({\it linear dispersion} = 0.72 \AA/pix) and in April 2007 with a
CCD mosaic optimised for the blue wavelengths. At my knowledge,
no stellar magnetic fields have been yet reported in the literature 
with this new set-up that is expected to improve the FORS1 capability
to measure stellar magnetic fields almost of a factor two at 4000 \AA. 

In this paper results of an attempt to measure stellar magnetic fields
with the {\it William Herschel Telescope}
equipped with the spectropolarimeter ISIS are reported. For a direct
comparison with FORS1 results, the blue region was selected.
In Section \ref{obs} observation strategy, data reduction and method
adopted to extract the polarisation signal 
are described. Section \ref{results} reports the results for well known
magnetic stars and a comparison with literature data obtained at
middle/high resolution circular spectropolarimetry. These literature
measurements can be considered as approaching values for an increasing
spectral resolution. The last Section \ref{concl}
summarises the capability of ISIS to measure stellar magnetic fields.

\section{Observations and data reduction}\label{obs}

To check the capability to measure stellar magnetic fields with ISIS at
the WHT the magnetic chemically peculiar stars: HD\,10783 (B=6.5 mag.), HD\,74521 (B=5.5 mag.) and HD\,201601 (B=5.0 mag.) have been observed. Moreover,
one of the stars with the weakest well established magnetic field has been observed: HD\,3360 (B=3.5 mag.), a perfect
target to determine any shift in the zero point of used set-up.

\begin{figure*}
\begin{center}
\includegraphics[width=14.5cm,height=5.5cm]{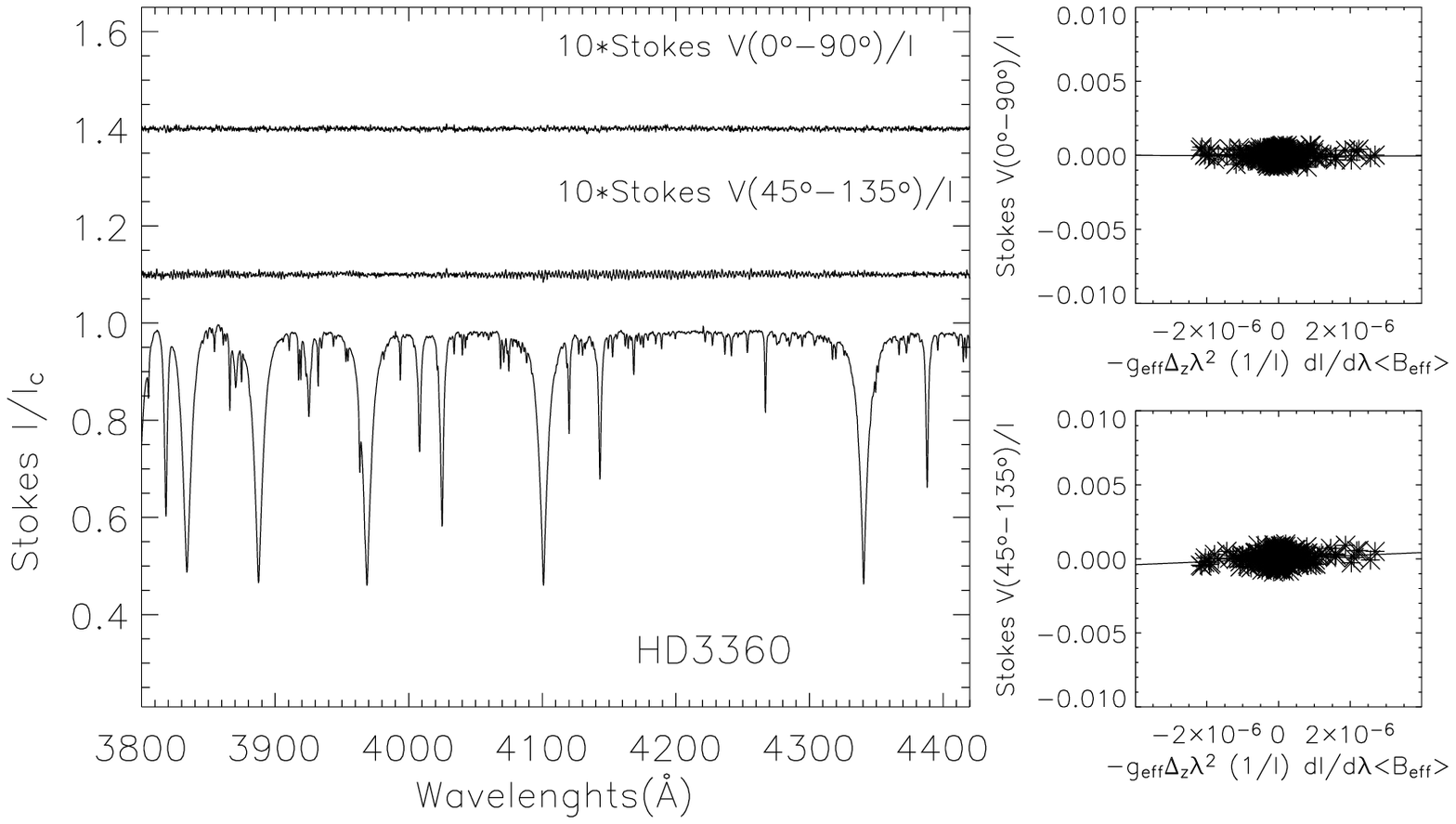}
\includegraphics[width=14.5cm,height=5.5cm]{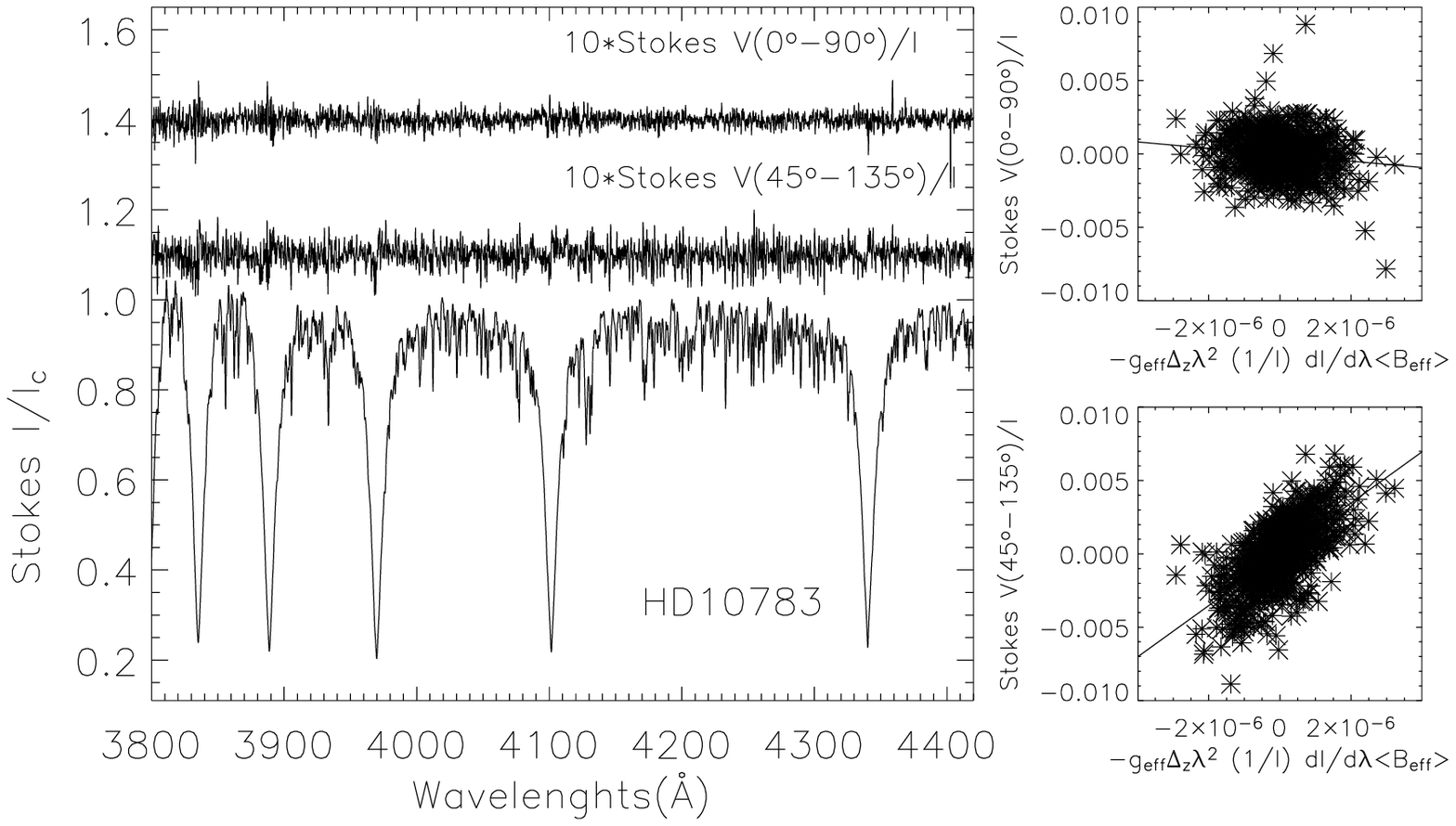}
\includegraphics[width=14.5cm,height=5.5cm]{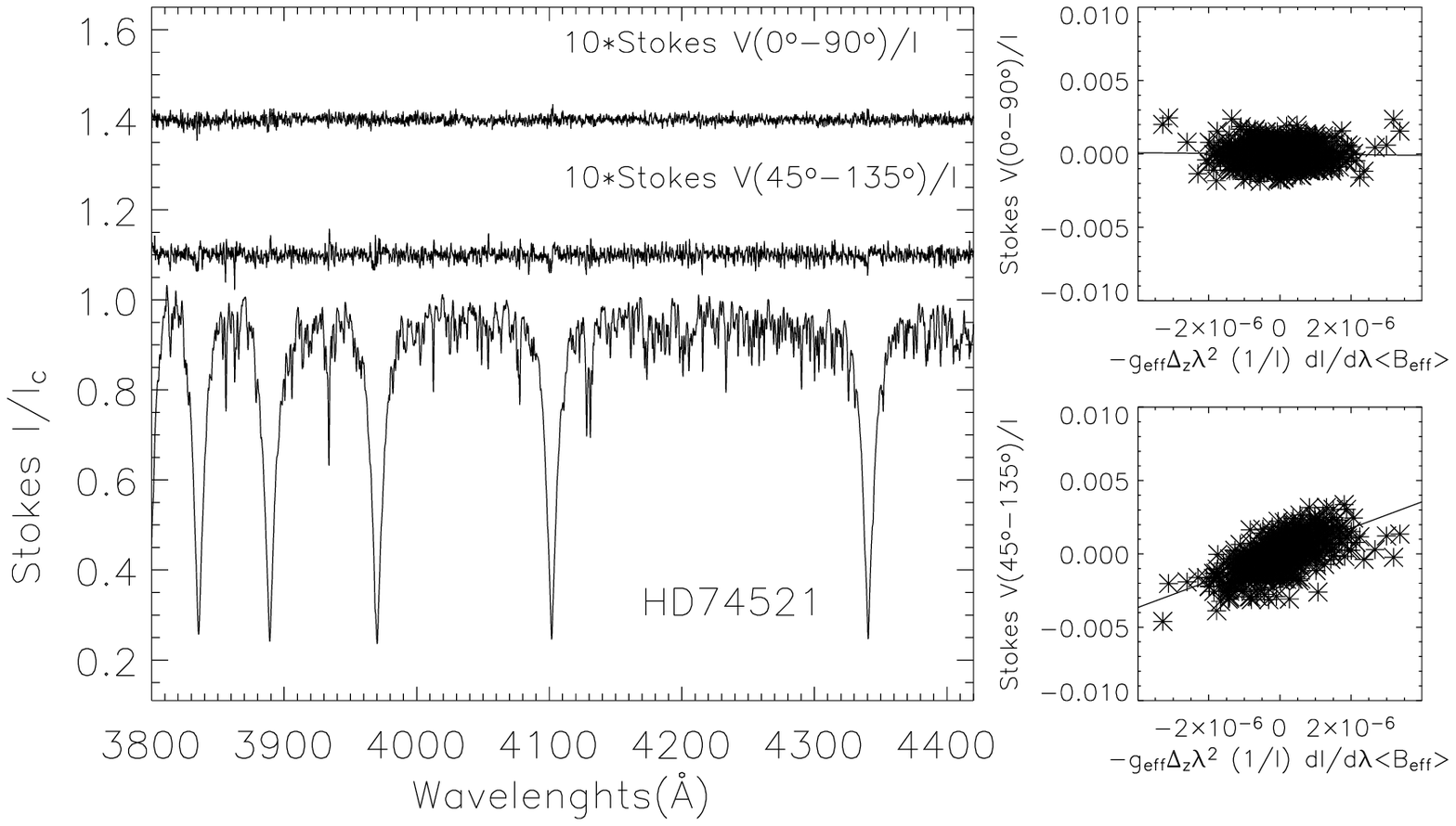}
\includegraphics[width=14.5cm,height=5.5cm]{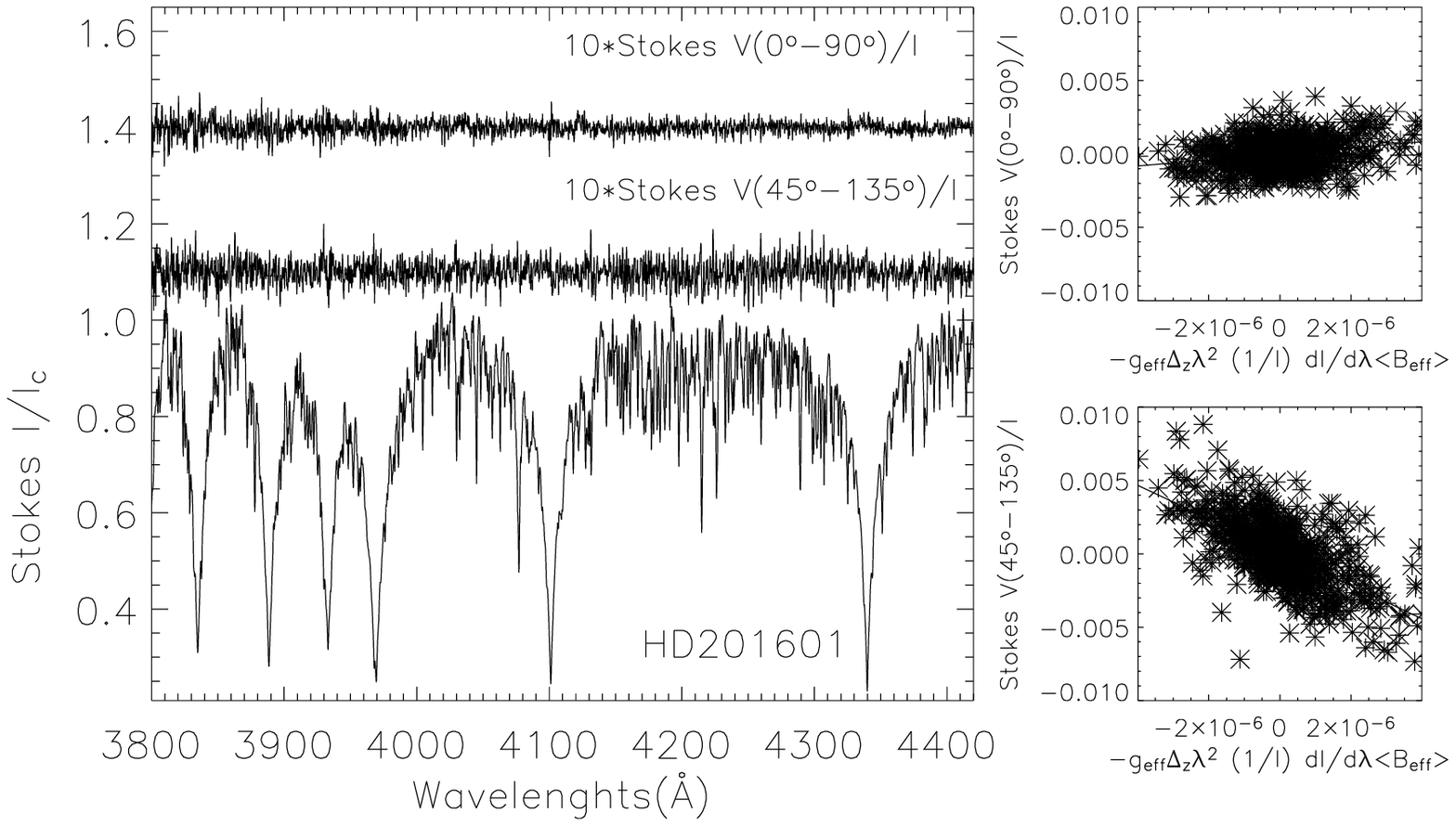}
\caption[]{Left panels show the Stokes $I$ and $V/I$ spectra for observed stars.
Stokes $V/I$ spectra are multiplied by ten and shifted for clearness.
Right panels show Stokes $V/I$ as function of $-g_{\rm eff} \ \Delta \lambda_z \ \lambda^{2} \ (1/I) \ {\rm d}I/{\rm d}\lambda$ 
adopting an {\it average} effective Land\'e factor $<$g$_{\rm eff}$$> = 1.0$.
Ripples are clearly present in the Stokes $V$ spectra of HD\,3360.}
\label{fig_spectra}
\end{center}
\end{figure*}

\begin{figure*}
\begin{center}
\hbox{
\includegraphics[width=8.4cm,height=4.5cm]{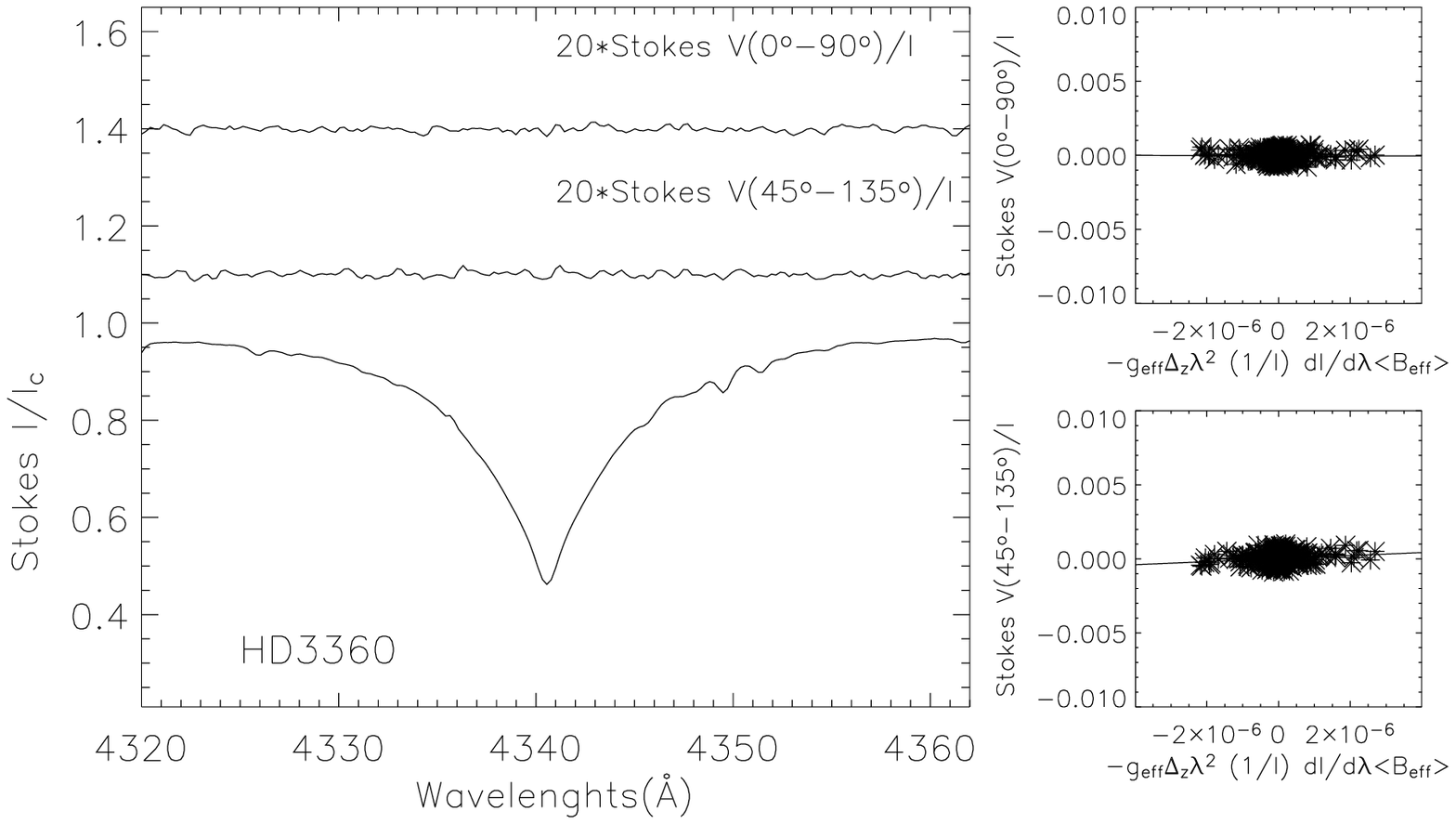}
\includegraphics[width=8.4cm,height=4.5cm]{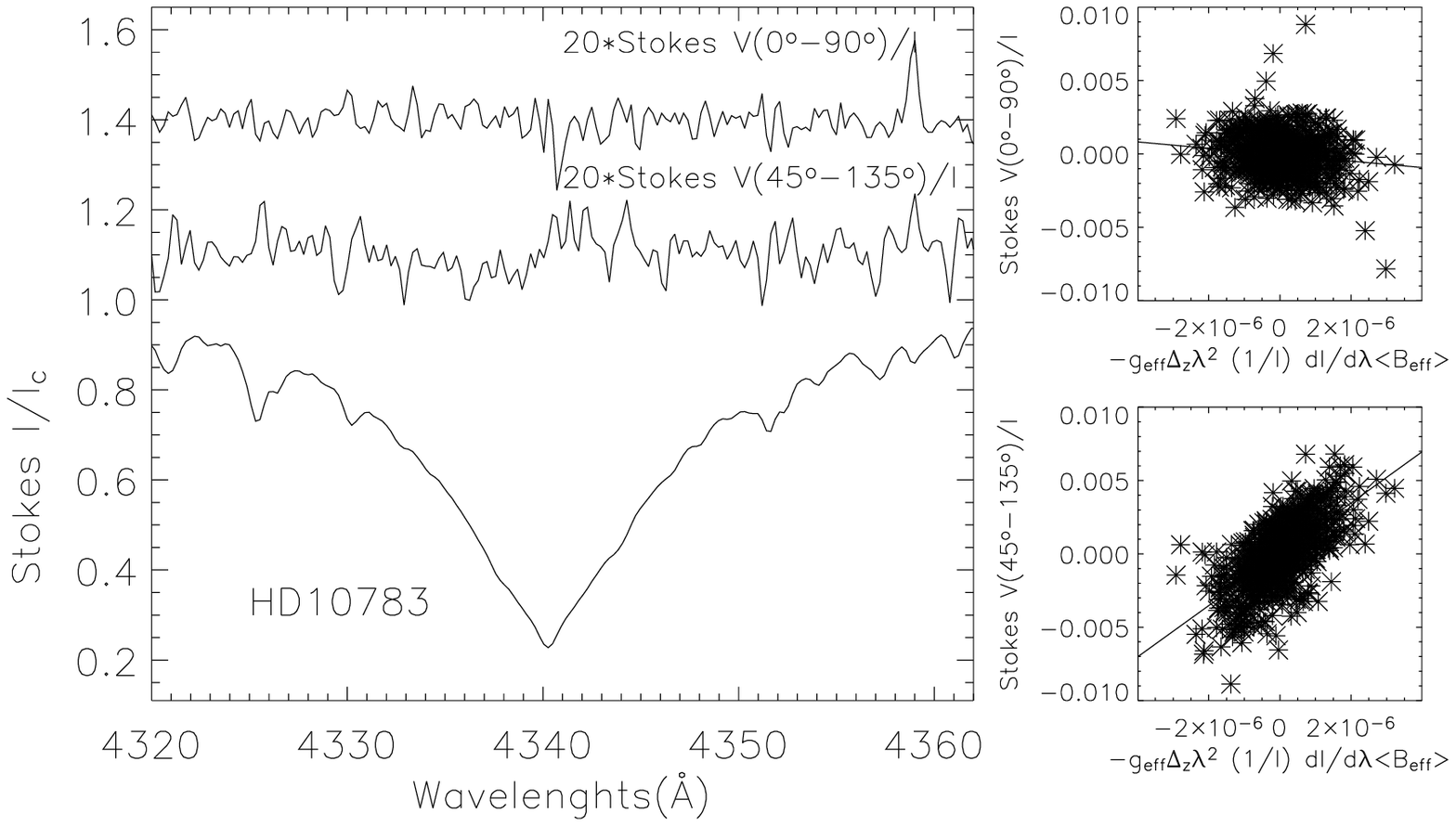}
                               }
\hbox{
\includegraphics[width=8.4cm,height=4.5cm]{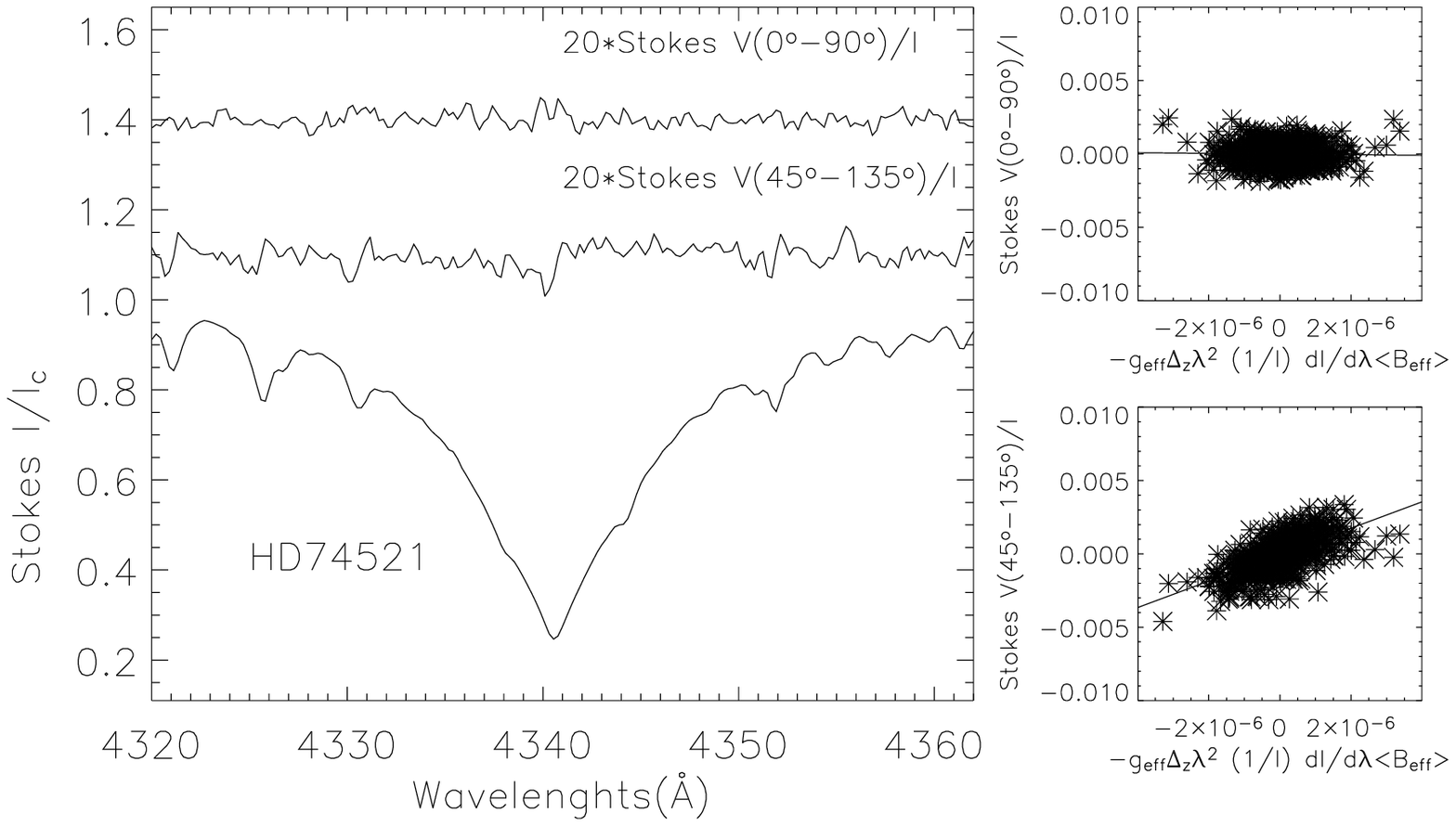}
\includegraphics[width=8.4cm,height=4.5cm]{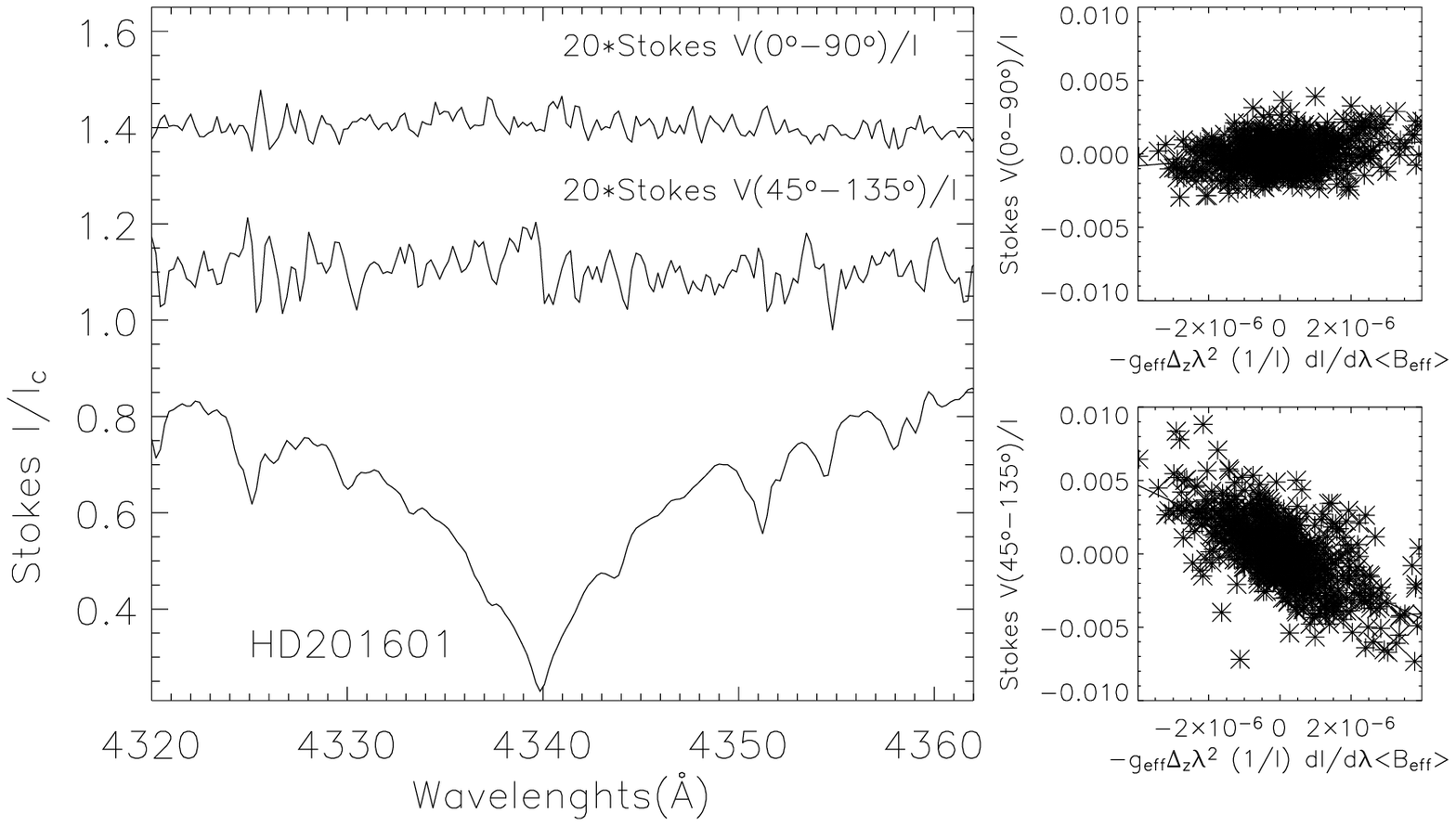}
                               }
\caption[]{Observed H$_{\gamma}$ profiles and their circular polarisation are plotted as in Fig. \ref{fig_spectra}, but Stokes $V/I$ spectra are multiplied by twenty.}
\label{fig_gamma}
\end{center}
\end{figure*}

On 2005 December 13 and 14 and on 2006 October 1, 2, 3 and 4,
circular spectropolarimetric observations in the 3785$-$4480 \AA\,
interval were carried out with the echelle grating R1200B at R=5000. Data
were reduced by using standard procedures for spectroscopic observations
within the NOAO/IRAF package as in Leone \& Kurtz \cite{Leone03}.

In general, a polarimeter consists of a {\it $\lambda/4$ retarder} and a {\it
polariser}. As a function of the angle $\alpha$ between the fast axis
of an ideal retarder and polariser acceptance axis, the circularly polarised
light spectrum - through the definition of the Stokes parameters - is \cite{Landi92}:
\begin{equation}
S(\alpha)=0.5[I-V\sin 2\alpha]
\end{equation}

The combination of the {\it o}-rdinary and {\it e}-xtraordinary
beams emerging from the polariser to measure the circular polarisation
degree is critical.  There is a time independent (instrumental)
sensitivity $G$, due to a pixel-by-pixel efficiency --
together with a time dependent sensitivity $F$ of spectra  obtained at
different $\alpha$ angles -- due to variation of sky
transparency and slit illumination. So that, photon noise dominated
Stokes $I$ and $V$
can been obtained from the recorded spectra at $\alpha$ = 45 and 135$^{\rm o}$:
\begin{eqnarray}
S_{45^{\rm o},o} = & 0.5\,(I + V) G_{o} F_{45^{\rm o}} \nonumber \\
S_{45^{\rm o},e} = & 0.5\,(I - V) G_{e} F_{45^{\rm o}} \nonumber \\
 &  \\
S_{135^{\rm o},o} = & 0.5\,(I - V) G_{o} F_{135^{\rm o}}\nonumber \\
S_{135^{\rm o},e} = & 0.5\,(I + V) G_{e} F_{135^{\rm o}} \nonumber
\end{eqnarray}
Hence:
\begin{equation}
\frac{V}{I} = \frac{R - 1}{R + 1}\hspace{1.2cm}{\rm
with}\hspace{0.6cm}R^2 = \frac{S_{45^{\rm o},o}/S_{45^{\rm o},e}}{S_{135^{\rm o},o}/S_{135^{\rm o},e}}
\end{equation}

For a real retarder the fast angle depends on wavelength and 
retarder and polariser are exactly aligned only at few
wavelengths. Eight spectra from the four exposures at $\alpha$ = 0, 45,
90 and 135$^{\rm o}$ give the possibility to obtain Stokes $I$ and
$V$ despite the unknown difference $\epsilon_\lambda$ from the real angle
between the retarder and polariser. In fact, replacing $\alpha$
with $\alpha+\epsilon_\lambda$ in eq. (3):
\begin{eqnarray}
S(90^{\rm o} + \epsilon_\lambda) - S(0^{\rm o} + \epsilon_\lambda) & = & V\sin 2\epsilon_\lambda \label{S0_90}\\
S(135^{\rm o} + \epsilon_\lambda) - S(45^{\rm o} + \epsilon_\lambda) & = & V\cos 2\epsilon_\lambda \label{S45_135}
\end{eqnarray}
For an ideal device, $\epsilon_\lambda$ is null and no circular polarisation
is measured through $S_{0^{\rm o}}$ and $S_{90^{\rm o}}$ spectra.
In practice, this is the case of ISIS polarimeter presenting no clear dependence
of $\epsilon_\lambda$ with wavelength. 
Fig.\,\ref{fig_spectra} shows the Stokes $I$ and $V/I$ spectra
for the entire interval and Fig.\,\ref{fig_gamma} shows the H$_{\gamma}$
region. Linear regressions of Stokes $V/I$ spectra as a function
of $-g_{\rm eff} \ \Delta \lambda_z \ \lambda^{2} \ (1/I) \ {\rm d}I/{\rm d}\lambda$ give, within errors, a null slope for $V\cos 2\epsilon_\lambda$.
A result consistent with the expectation that no circular polarised flux
is measured through Stokes $V$($S_{0^{\rm o}},S_{90^{\rm o}}$) spectra.

\begin{table}
\caption{Observed stars. Effective magnetic fields
have been measured from eq. (\ref{eqV}) assuming
an {\it average} effective Land\'e factor $<$g$_{\rm eff}$$> = 1.0$
across the whole spectrum and through Balmer lines. Heliocentric Julian Data($-$2\,400\,000) and
S/N ratios measured on Stokes $I/V$ continuum are reported. }
\label{Tab_Standard}
      \[
\scriptstyle
         \begin{array}{crrrr}   
            \noalign{\smallskip}\hline
\multicolumn{5}{c}{\rm HD\,3360}\\
\multicolumn{2}{r}{\rm HJD =  53718.323} & & \\
\multicolumn{2}{r}{S/N = ~~~~~~3300}&\multicolumn{1}{c}{}&\multicolumn{1}{c}{}\\
             &  \multicolumn{1}{r}{B_{\rm eff}~~~~}   & &  & \\
\hline
    All        &     78\pm ~~20  & &  & \\
                       &               & &  & \\
H_{\rm  9}  &     183\pm 104 & &  & \\
H_{\rm  8}  &     134\pm ~~73 & &  & \\
H_{\epsilon}&      98\pm ~~59 & &  & \\
H_{\delta}  &      18\pm ~~85 & &  & \\
H_{\gamma}  &     121\pm ~~50 & &  & \\
 Av. &114\pm ~~67 & &  &  \\
\hline
            \noalign{\smallskip}\hline
\multicolumn{5}{c}{\rm HD\,10783}\\
\multicolumn{2}{r}{\rm HJD = 54010.542}&\multicolumn{1}{r}{54011.620}&\multicolumn{1}{r}{54012.597}&\multicolumn{1}{r}{54013.575}\\
\multicolumn{2}{r}{S/N = ~~~~~~~~140}&\multicolumn{1}{r}{600}&\multicolumn{1}{r}{890}&\multicolumn{1}{r}{230}\\
     & \multicolumn{1}{r}{B_{\rm eff}~~~~} &  \multicolumn{1}{r}{B_{\rm eff}~~~~} & \multicolumn{1}{r}{B_{\rm eff}~~~~} &  \multicolumn{1}{r}{B_{\rm eff}~~~~} \\\hline
 All &   1795 \pm ~~330 & 1921 \pm ~46&  309 \pm   ~~42 & -426 \pm ~~136\\
                       &             &            & & \\
H_{\rm  9}   &  -1329 \pm 1450  & 1826 \pm 199&  259 \pm  160 &  1311 \pm  ~~550 \\
H_{\rm  8}   &   3753 \pm 1245  & 1668 \pm 168&  433 \pm  146 & -1934 \pm  ~~606 \\
H_{\epsilon} &  -1086 \pm 1058  & 1680 \pm 151&  279 \pm  113 &  -374 \pm  ~~502 \\
H_{\delta}   &   3751 \pm ~~989 & 1764 \pm 133&   -6 \pm  110 & -1596 \pm  ~~431 \\
H_{\gamma}   &   3285 \pm ~~851 & 1447 \pm 120&  226 \pm  109 &  -940 \pm  ~~307 \\
   Av.       &   1887 \pm 2563  & 1660 \pm 150&  251 \pm  156 & -674 \pm 1371  \\  
\hline
            \noalign{\smallskip}\hline
\multicolumn{5}{c}{\rm HD\,74521}\\
\multicolumn{2}{r}{\rm HJD = 53718.716}&\multicolumn{1}{r}{53719.709}&\multicolumn{1}{r}{}&\multicolumn{1}{c}{}\\
\multicolumn{2}{r}{S/N = ~~~~~~1300}&\multicolumn{1}{r}{1100}& &\\
     & \multicolumn{1}{r}{B_{\rm eff}~~~~} &  \multicolumn{1}{r}{B_{\rm eff}~~~~} \\\hline
    All        & 1032\pm ~~30 & 794\pm  28 & & \\
                       &             &            & & \\
H_{\rm  9}  &  979\pm 137  & 556\pm 138  & &  \\
H_{\rm  8}  &  831\pm 115  & 459\pm 117  & &  \\
H_{\epsilon}&  900\pm 109  & 876\pm 101  & &  \\
H_{\delta}  &  881\pm 108  & 793\pm 122  & &  \\
H_{\gamma}  &  808\pm ~~73 & 765\pm ~~72 & &  \\
 Av. &  889\pm ~~68 & 706\pm 169 & &  \\
\hline
            \noalign{\smallskip}\hline
\multicolumn{5}{c}{\rm HD\,201601}\\
\multicolumn{2}{r}{\rm HJD = 54010.428}&\multicolumn{1}{r}{54011.620}\\
\multicolumn{2}{r}{S/N = ~~~~~~~~230}&\multicolumn{1}{r}{860}\\
     & \multicolumn{1}{r}{B_{\rm eff}~~~~} &  \multicolumn{1}{r}{B_{\rm eff}~~~~} &  \multicolumn{1}{r}{~~~~}\\\hline
    All      &  -1143\pm 113  & -1114\pm ~~34 & &\\
             &                &               & &\\
 H_{\rm  9}  &  -1148\pm 566  & -754\pm 196   & &\\
 H_{\rm  8}  &   -996\pm 570  &  -90\pm 154   & &\\
 H_{\epsilon}&    603\pm 571  & -730\pm 132   & &\\
 H_{\delta}  &  -1694\pm 438  &-1098\pm 106   & &\\
 H_{\gamma}  &   -648\pm 298  &-1109\pm ~~73  & &\\
 Av.         &   -802\pm 811  & -839\pm 406   & &\\
\hline
            \noalign{\smallskip}
         \end{array}
      \]
   \end{table}
 
\section{Measurement of the effective magnetic field at low resolution}\label{results}

Observing a star at high spectral resolution, it is possible to select
unblended metal lines and, according to their effective Land\`e factor,
circular polarisation is converted in a measure of the effective magnetic field
through eq. (\ref{eqRV}).
A few remarks are necessary to understand the effective magnetic fields
measured at low resolution by means of eq. (\ref{eqV}) when an hardware
average of Stokes $I$ and $V$ in wavelength is performed:
\begin{itemize}
\item by means of R = 115\,000 circular spectropolarimetry,
Leone \& Catanzaro \cite{Leone04} and Leone \& Kurtz \cite{Leone03} have shown that the measured effective
magnetic field of the chemically peculiar star HD\,24712 and HD\,201601 is a
function
of the selected element with values increasing with the atomic number.
The most straight explanation, strengthened by the variable equivalent widths
with the magnetic period, is a not homogeneous distribution of elements
on the stellar surface with a concentration of heavy elements in the magnetic
polar regions. These stars show that {\it at priori} there is no justification in
combining magnetic field measurements obtained from different spectral lines.
At low spectral resolution any information on the spatial and vertical
distribution of the magnetic fields can be lost and the field under estimated.
\item absorption and emission lines present opposite circular polarisation
if formed in the same magnetic region. A difference in sign of Stokes $V$
that led to an under estimation of the magnetic field of stars characterised by
a chromosphere or an envelope. 
\item an {\it average} effective Land\'e factor ($<$g$_{\rm eff}$$>$) is
necessary to correctly
convert the measured polarisation in an effective magnetic field on the basis
of eq. (\ref{eqV}). This depends on spectral type, gravity and chemical
composition of the observed star. Moreover because of the instrumental smoothing,
$<$g$_{\rm eff}$$>$ depends on the resolution of the spectropolarimeter too.
A numerical simulation, for the used observational set-up, has been carried
out with COSSAM \cite{Stift00} for a main sequence A-type star with solar
abundances and homogeneous distribution of elements in the photosphere,
the average Land\'e factors of the metal lines present in the
3800$-$4400 \AA\, interval is $<$g$_{\rm eff}$$>$ = 1.1. 
A correct evaluation of the average effective Land\'e factor for a star
depends on the relative strength of its spectral lines and it requires 
the knowledge of atmosphere parameters with particular attention to
abundances. As to the here observed chemically peculiar stars,
abundances cannot be here derived at low resolution and
$<$g$_{\rm eff}$$>$ = 1.0 was adopted in this exploratory study.
\end{itemize}

\subsection{HD\,3360}
The first attempt to measure the effective magnetic field of the slow pulsating
B-type star HD\,3360 was due to Landstreet \cite{Landstreet82}, who measured a field
equal to 70$\pm$310 G. Neiner et al. \cite{Neiner03} 
discovered a 5\fd370447 periodic effective magnetic field with an average
value of $-$18 G and a 30 G amplitude. 

Circular spectropolarimetry of HD\,3360 has been carried on on Dec. 13, 2005
when, according to Neiner's ephemeris, a 10 G
field was expected.  Stokes $I$ and Stokes $V/I$ spectra are shown in
Fig.\,\ref{fig_spectra} and Fig.\,\ref{fig_gamma}. A S/N = 3300 is measured in intervals 
free of spectral lines of the Stokes $V/I$ spectra. 

To measure the effective magnetic field of HD\,3360 through eq.\ (\ref{eqV})
a least square fits of V($S_{45^{\rm o}},S_{135^{\rm o}}$) and V($S_{0^{\rm o}},S_{90^{\rm o}}$) vs.
$-g_{\rm eff} \ \Delta \lambda_z \ \lambda^{2} \ (1/I) \ {\rm d}I/{\rm d}\lambda$
have been performed. Despite within errors  V($S_{0^{\rm o}},S_{90^{\rm o}}$) is null,
to steady apply eq. (\ref{eqV}), 
B$_{\rm eff}$ was obtained adding quadratically eqs.(\ref{S0_90}) and (\ref{S45_135}).
Analogously, errors on the measured fields have been estimated.

A value of 78$\pm$20 G has been measured along the whole spectrum
for  $<$g$_{\rm eff}$$>$ = 1.0 (Table\,1). 
Certainly a more realistic Land\'e factor would get the measured field closer
to the literature value, however ripples are equally present in the Stokes
$V$($S_{45^{\rm o}},S_{135^{\rm o}}$) and $V$($S_{0^{\rm o}},S_{90^{\rm o}}$)
spectra of HD\,3360 (Fig.\,\ref{fig_spectra} and Fig.\,\ref{fig_gamma}).
Probably due to interferences in the retarder, these ripples result in a
instrumental systematic error not larger than 80 G.

\subsection{HD\,201601}
This very famous magnetic chemically peculiar star HD\,201601 shows an effective
magnetic field that 
is sinusoidal variable between $-262\pm$839 G with a period of $91.1\pm$3.6 years \cite{Bychkov06}.

HD\,201601 was observed on Oct. 1 and 2, 2006. Fig.\,\ref{fig_spectra} shows
the crowded spectrum of this F-type metal rich star,
while the Fig.\,\ref{fig_gamma} shows the H$_{\gamma}$ profile, obtained in the second night.

Differently than HD\,3360, the slope of Stokes $V$ versus
$-g_{\rm eff} \ \Delta \lambda_z \ \lambda^{2} \ (1/I) \ {\rm d}I/{\rm d}\lambda$
testifies the strong circular polarisation degree of HD\,201601 spectrum.
Converted in an effective magnetic field for $<$g$_{\rm eff}$$>$ = 1.0, 
slopes give $-1143\pm 113$ and $-1114\pm 34$ G for the two consecutive nights respectively (Table\ 1).
According to Bychkov's ephemeris, the ISIS observations of HD\,201601 were
carried out very close to the magnetic minimum  $\rm B_{\rm eff} = -1101\pm 3 1$ G. It appears that within errors,
the here measured field of HD\,201601 is coincident to the literature value. 

Table\,1 shows that for this F-type star characterised by large metal
overabundances a truthful measure of the effective field can
be obtained using the whole spectrum as soon as S/N$\sim$250. Single Balmer lines are indicative
of the magnetic field strength only from S/N$\ge$850.

\subsection*{HD~10783 = GC~2141 = UZ~Psc}
As normal for a magnetic chemically peculiar star \cite{Wolff83}, 
HD\,10783 is expected to present photometric and magnetic variability with the rotational
period.  

Photometric UBV observations of this star were obtained by Hardie et 
al. \cite{Hardie90}, who derived a period $P = $4\fd13282$\pm$0.00014 by  
fitting the variable data to the function:
\begin{eqnarray}
f = A_0 & + A_1 \sin 2\pi\left(\frac{(t-t_0)}{P}+\phi_1\right) \nonumber \\
                    & + A_2 \sin 2\pi\left(\frac{2(t-t_0)}{P}+\phi_2\right) \label{eqsin}
\end{eqnarray} 
where $t$ is the time of observations and $t_0$ is a reference instant,

Following Leone \& Catanzaro \cite{Leone01}, {\it Hipparcos} photometry \cite{ESA97} is combined with all available data in the literature by Preston \& Stepien \cite{Preston68}, van Genderen
\cite{VanGenderen67}. Any single data set has been fit to
Hardie et al. \cite{Hardie90} function to determine and remove systematic differences in brightness ($A_0$). Then a least square fit of
whole data gives the ephemeris:
$$
JD(\rm max) = 2\,439\,757.88\pm0.11 + 4.13275\pm 0.00009
$$

\begin{figure}
\begin{center}
\includegraphics[width=8.5cm]{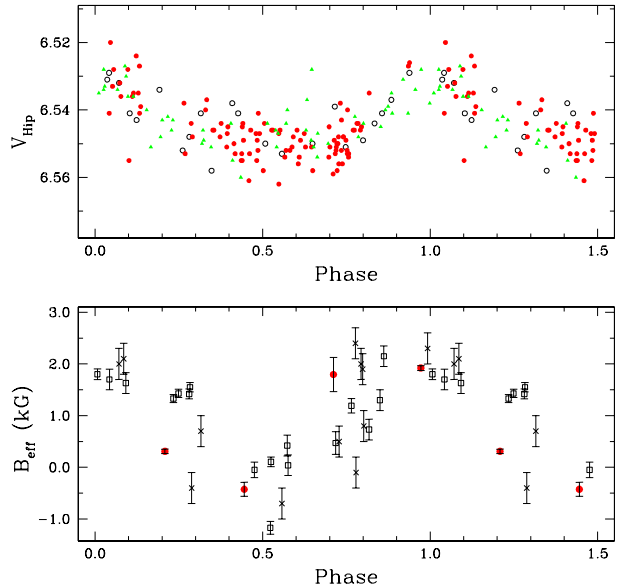}
\caption[]{Photometric and magnetic variability of HD\,10783 folded for
the here determined $4\fd13275$ period. Top panel shows photometric data by Preston \& Stepien \cite{Preston68} (empty circles),
van Genderen \cite{VanGenderen67} (filled triangles) and {\it Hipparcos} satellite \cite{ESA97} (filled circles). The bottom panel shows measurements
of the effective magnetic field by Babcock \cite{Babcock58} (empty squares) and Preston \& Stepien \cite{Preston68} (crosses).
Circles represent here presented low resolution measurements for
$<$g$_{\rm eff}$$>$ = 1.0.  Error bars extend to twice the estimated error.}
\label{fig_HD10783_var}
\end{center}
\end{figure}

The top panel of Fig.\,\ref{fig_HD10783_var} shows the photometric variability
of HD\,10783 folded with the here determined ephemeris.

HD\,10783 was observed on Oct. 1, 2, 3 and 4, 2006 particularly with the aim
to check the capability of ISIS to monitor a variable magnetic field.
The observed spectrum of HD\,10783 obtained on October 2, 2006
is entirely plotted in Fig.\,\ref{fig_spectra}, while the $H_{\gamma}$ is
in Fig.\,\ref{fig_gamma}. Measurements of the effective magnetic field
on the basis of eq. (\ref{eqV}), for $<$g$_{\rm eff}$$>$ = 1.0, are listed
in Table\,1. 

A comparison with the literature data shows (bottom panel of
Fig.\ \ref{fig_HD10783_var}) that $\rm B_{eff}$ measures are
as variable as expected. These field values are indicative
of the period 4\fd13267, that is - within error - slightly shorter 
than the photometric period.
A result, based on magnetic data spanning 20523 days instead of 
the 10619 days covered with the photometric data,
to be confirmed by further {\it classical} measurements of
the effective magnetic field  of HD\,10783.

\subsection{HD\,74521}
The most recent determination of the variability period of the magnetic chemically peculiar star HD\,74521 has been
obtained by Adelman \cite{Adelman98}, who found 7\fd0501\,d. 

\begin{figure}
\begin{center}
\includegraphics[width=8.5cm]{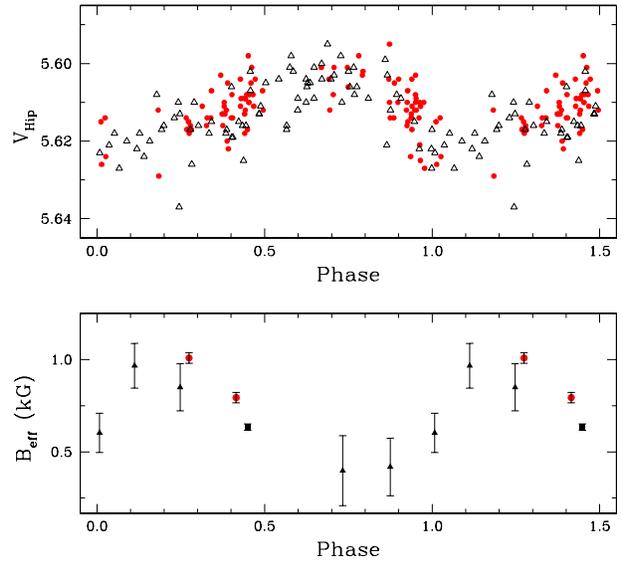}
\caption[]{Photometric and magnetic variability of HD\,74521 assuming
the here determined $7\fd0486$  period.  Top panel shows Hipparcos satellite photometry \cite{ESA97} (circles)
and Adelman \cite{Adelman98} photometry in the $b$ Str\"omgren filter (triangles).
The bottom panel shows the effective magnetic field as measured by Mathys
\cite{Mathys94} (triangles) and Shorlin et al. \cite{Shorlin02} (square).
Circles represent here presented low resolution measurements for
$<$g$_{\rm eff}$$>$ = 1.0. Error bars extend to twice the estimated error.}
\label{fig_HD74521_var}
\end{center}
\end{figure}

Catalano \& Leone \cite{Catalano93} noted that HD\,74521 light curves are asymmetric with the rise to maximum
taking longer than the fall to minimum. For this reason a fit based on eq. (\ref{eqsin}) is not appropriate and a period
search applying the phase dispersion minimisation method of Stellingwerf \cite{Stellingwerf78} to Adelman \cite{Adelman98} and {\it Hipparcos} \cite{ESA97} data has been performed. The most probable period is $P= 7\fd0486\pm$0.005, whose error is estimated by the width of the found minimum. Folded photometric data are in 
the top panel of Fig.\,\ref{fig_HD74521_var}.

The effective magnetic field of HD\,74521 has been measured,
for $<$g$_{\rm eff}$$>$ = 1.0, from spectra collected
on Dec. 13 and 14, 2005. These high S/N spectra state
that the measures within Balmer lines are about 15\% smaller
than from the whole spectrum (Table\ 1). A result presents, within errors,
also in the HD\,10783 and HD\,201601 measurements. 
A difference probably due to the adopted Land\'e factors.
It has to be noted that for a Balmer line g$_{\rm eff}$ = 1.0
while for the metal lines COSSAM ascribed to an A-type star
$<$g$_{\rm eff}$$>$ = 1.1.

Adopting the here obtained variability period, the low resolution ISIS measures of HD\,74521 are compared to
the middle resolution spectropolarimetric values obtained by Mathys \cite{Mathys94} and Shorlin et al. \cite{Shorlin02}
(Fig.~\ref{fig_HD74521_var}). 
The effective field measured at low resolution is slightly larger
than these literature values, probably indicating the effect of the
80 G instrumental error and/or that $<$g$_{\rm eff}$$>$ = 1.1 is
more appropriate for the A-type star HD\,74521.

\section{Conclusion}\label{concl}
Circular low resolution (R$=$5000) spectropolarimetric observations
have been carried out with ISIS spectrograph at the 4.5\,m WHT in
the 3785$-$4480 \AA\ interval of three magnetic chemically
peculiar stars and HD\,3360, one of the stars with the weakest
well documented magnetic field.

 Measured effective magnetic fields
on the basis of eq. (\ref{eqV}) assuming an average Land\'e factor $<$g$_{\rm eff}$$> = 1.0$
are in a general agreement with values reported in the literature.
Internal errors are: 330 G for S/N = 140,
140 G for S/N = 230 and $<$50 G for S/N$>$ 600.
Ripples are clearly present in the HD\,3360 spectra resulting in 
a (positive field) instrumental error smaller than 80 G. A S/N larger
than 1000 is necessary for 
a reliable estimation of the effective magnetic field from Balmer lines
that is, however, about 85\% of the value obtained from the
whole spectrum. 

In principle, exposures at four different $\alpha$ angles between
the retarder and polariser are necessary to correctly measure the circular
polarisation. As to ISIS polarimeter, no circular polarisation was measured
through
the spectra obtained at $\alpha = 0$ and $90^{\rm o}$, an important condition
that reduces from eight to four the number of spectra necessary to obtain
photon noise dominated Stokes $I$ and\,$V$.

The here presented results show that observational support to the
theoretical studies on stellar magnetism is not restricted to the overloaded
ten meter class telescopes. Widespread four
meter class telescopes can also detect and measure the magnetic
field of faint stars. As an example, for the here adopted
set-up in average observing condition (seeing = 0.8$\arcsec$ and airmass = 1.3)
ISIS exposure calculator predicts an exposure time of one hour to measure
the magnetic field of a B = 12 mag. F-A type star with a precision
of 300 G. on the contrary, the very high S/N necessary to measure the effective magnetic field
from Balmer lines only is a significative limit for the hottest stars.

\end{document}